\documentclass[prl,twocolumn,floatfix,showpacs,preprintnumbers,amsmath,amssymb,superscriptaddress,nofootinbib]{revtex4}

\usepackage{amsmath}
\usepackage{amssymb}
\usepackage{amsthm}
\usepackage{overpic,color}
\usepackage{afterpage} 
\usepackage{epsfig}
\usepackage{float}
\usepackage{rotating}
\usepackage{lscape}
\usepackage{enumerate}
\usepackage{xspace}
\usepackage{url}
\usepackage{upgreek}
\usepackage{dcolumn}      
\usepackage{bm}           
\usepackage{multirow}
\usepackage{bigstrut}
\usepackage{upgreek}
\usepackage[tight]{units} 

\newcommand{\numu}{\mbox{$\nu_{\mu}$}}                   
\newcommand{\nue}{\mbox{$\nu_{e}$}}                      

\newcommand{\num}{\numu}

\newcommand{\nus}{\nu_s}
\newcommand{\nut}{\nu_{\tau}}

\begin{document}
\preprint{FERMILAB-PUB-11-183-E, BNL-95065-2011-JA, arXiv:hep-ex/1104.3922}
\title{Active to sterile neutrino mixing limits from neutral-current interactions in MINOS}         

\newcommand{\Berkeley}{Lawrence Berkeley National Laboratory, Berkeley, California, 94720 USA}
\newcommand{\Cambridge}{Cavendish Laboratory, University of Cambridge, Madingley Road, Cambridge CB3 0HE, United Kingdom}
\newcommand{\FNAL}{Fermi National Accelerator Laboratory, Batavia, Illinois 60510, USA}
\newcommand{\RAL}{Rutherford Appleton Laboratory, Science and Technologies Facilities Council, Didcot OX11 0QX, United Kingdom}
\newcommand{\UCL}{Department of Physics and Astronomy, University College London, Gower Street, London WC1E 6BT, United Kingdom}
\newcommand{\Caltech}{Lauritsen Laboratory, California Institute of Technology, Pasadena, California 91125, USA}
\newcommand{\Alabama}{Department of Physics and Astronomy, University of Alabama, Tuscaloosa, Alabama 35487, USA}
\newcommand{\ANL}{Argonne National Laboratory, Argonne, Illinois 60439, USA}
\newcommand{\Athens}{Department of Physics, University of Athens, GR-15771 Athens, Greece}
\newcommand{\NTUAthens}{Department of Physics, National Tech. University of Athens, GR-15780 Athens, Greece}
\newcommand{\Benedictine}{Physics Department, Benedictine University, Lisle, Illinois 60532, USA}
\newcommand{\BNL}{Brookhaven National Laboratory, Upton, New York 11973, USA}
\newcommand{\CdF}{APC -- Universit\'{e} Paris 7 Denis Diderot, 10, rue Alice Domon et L\'{e}onie Duquet, F-75205 Paris Cedex 13, France}
\newcommand{\Cleveland}{Cleveland Clinic, Cleveland, Ohio 44195, USA}
\newcommand{\Delhi}{Department of Physics \& Astrophysics, University of Delhi, Delhi 110007, India}
\newcommand{\GEHealth}{GE Healthcare, Florence South Carolina 29501, USA}
\newcommand{\Harvard}{Department of Physics, Harvard University, Cambridge, Massachusetts 02138, USA}
\newcommand{\HolyCross}{Holy Cross College, Notre Dame, Indiana 46556, USA}
\newcommand{\IIT}{Physics Division, Illinois Institute of Technology, Chicago, Illinois 60616, USA}
\newcommand{\Iowa}{Department of Physics and Astronomy, Iowa State University, Ames, Iowa 50011 USA}
\newcommand{\Indiana}{Indiana University, Bloomington, Indiana 47405, USA}
\newcommand{\ITEP}{High Energy Experimental Physics Department, ITEP, B. Cheremushkinskaya, 25, 117218 Moscow, Russia}
\newcommand{\JMU}{Physics Department, James Madison University, Harrisonburg, Virginia 22807, USA}
\newcommand{\LASL}{Nuclear Nonproliferation Division, Threat Reduction Directorate, Los Alamos National Laboratory, Los Alamos, New Mexico 87545, USA}
\newcommand{\Lebedev}{Nuclear Physics Department, Lebedev Physical Institute, Leninsky Prospect 53, 119991 Moscow, Russia}
\newcommand{\LLL}{Lawrence Livermore National Laboratory, Livermore, California 94550, USA}
\newcommand{\LosAlamos}{Los Alamos National Laboratory, Los Alamos, New Mexico 87545, USA}
\newcommand{\MIT}{Lincoln Laboratory, Massachusetts Institute of Technology, Lexington, Massachusetts 02420, USA}
\newcommand{\Minnesota}{University of Minnesota, Minneapolis, Minnesota 55455, USA}
\newcommand{\Crookston}{Math, Science and Technology Department, University of Minnesota -- Crookston, Crookston, Minnesota 56716, USA}
\newcommand{\Duluth}{Department of Physics, University of Minnesota -- Duluth, Duluth, Minnesota 55812, USA}
\newcommand{\Ohio}{Center for Cosmology and Astro Particle Physics, Ohio State University, Columbus, Ohio 43210 USA}
\newcommand{\Otterbein}{Otterbein College, Westerville, Ohio 43081, USA}
\newcommand{\Oxford}{Subdepartment of Particle Physics, University of Oxford, Oxford OX1 3RH, United Kingdom}
\newcommand{\PennState}{Department of Physics, Pennsylvania State University, State College, Pennsylvania 16802, USA}
\newcommand{\PennU}{Department of Physics and Astronomy, University of Pennsylvania, Philadelphia, Pennsylvania 19104, USA}
\newcommand{\Pittsburgh}{Department of Physics and Astronomy, University of Pittsburgh, Pittsburgh, Pennsylvania 15260, USA}
\newcommand{\IHEP}{Institute for High Energy Physics, Protvino, Moscow Region RU-140284, Russia}
\newcommand{\Rochester}{Department of Physics and Astronomy, University of Rochester, New York 14627 USA}
\newcommand{\RoyalH}{Physics Department, Royal Holloway, University of London, Egham, Surrey, TW20 0EX, United Kingdom}
\newcommand{\Carolina}{Department of Physics and Astronomy, University of South Carolina, Columbia, South Carolina 29208, USA}
\newcommand{\SLAC}{Stanford Linear Accelerator Center, Stanford, California 94309, USA}
\newcommand{\Stanford}{Department of Physics, Stanford University, Stanford, California 94305, USA}
\newcommand{\StJohnFisher}{Physics Department, St. John Fisher College, Rochester, New York 14618 USA}
\newcommand{\Sussex}{Department of Physics and Astronomy, University of Sussex, Falmer, Brighton BN1 9QH, United Kingdom}
\newcommand{\TexasAM}{Physics Department, Texas A\&M University, College Station, Texas 77843, USA}
\newcommand{\Texas}{Department of Physics, University of Texas at Austin, 1 University Station C1600, Austin, Texas 78712, USA}
\newcommand{\TechX}{Tech-X Corporation, Boulder, Colorado 80303, USA}
\newcommand{\Tufts}{Physics Department, Tufts University, Medford, Massachusetts 02155, USA}
\newcommand{\UNICAMP}{Universidade Estadual de Campinas, IFGW-UNICAMP, CP 6165, 13083-970, Campinas, SP, Brazil}
\newcommand{\UFG}{Instituto de F\'{i}sica, Universidade Federal de Goi\'{a}s, CP 131, 74001-970, Goi\^{a}nia, GO, Brazil}
\newcommand{\USP}{Instituto de F\'{i}sica, Universidade de S\~{a}o Paulo,  CP 66318, 05315-970, S\~{a}o Paulo, SP, Brazil}
\newcommand{\Warsaw}{Department of Physics, Warsaw University, Ho\.{z}a 69, PL-00-681 Warsaw, Poland}
\newcommand{\Washington}{Physics Department, Western Washington University, Bellingham, Washington 98225, USA}
\newcommand{\WandM}{Department of Physics, College of William \& Mary, Williamsburg, Virginia 23187, USA}
\newcommand{\Wisconsin}{Physics Department, University of Wisconsin, Madison, Wisconsin 53706, USA}
\newcommand{\deceased}{Deceased.}

\affiliation{\ANL}
\affiliation{\Athens}
\affiliation{\BNL}
\affiliation{\Caltech}
\affiliation{\Cambridge}
\affiliation{\UNICAMP}
\affiliation{\FNAL}
\affiliation{\UFG}
\affiliation{\Harvard}
\affiliation{\HolyCross}
\affiliation{\IIT}
\affiliation{\Indiana}
\affiliation{\Iowa}
\affiliation{\UCL}
\affiliation{\Minnesota}
\affiliation{\Duluth}
\affiliation{\Otterbein}
\affiliation{\Oxford}
\affiliation{\Pittsburgh}
\affiliation{\RAL}
\affiliation{\USP}
\affiliation{\Carolina}
\affiliation{\Stanford}
\affiliation{\Sussex}
\affiliation{\TexasAM}
\affiliation{\Texas}
\affiliation{\Tufts}
\affiliation{\Warsaw}
\affiliation{\WandM}

\author{P.~Adamson}
\affiliation{\FNAL}

\author{D.~J.~Auty}
\affiliation{\Sussex}

\author{D.~S.~Ayres}
\affiliation{\ANL}

\author{C.~Backhouse}
\affiliation{\Oxford}

\author{G.~Barr}
\affiliation{\Oxford}

\author{M.~Bishai}
\affiliation{\BNL}

\author{A.~Blake}
\affiliation{\Cambridge}

\author{G.~J.~Bock}
\affiliation{\FNAL}

\author{D.~J.~Boehnlein}
\affiliation{\FNAL}

\author{D.~Bogert}
\affiliation{\FNAL}

% ...

\collaboration{The MINOS Collaboration}
\noaffiliation

\date{\today}
\begin{abstract}

Results are reported from a search for active to sterile neutrino oscillations in the MINOS long-baseline experiment, based on the observation of neutral-current neutrino interactions, from an exposure to the NuMI neutrino beam of $7.07\times10^{20}$~protons on target. A total of 802 neutral-current event candidates is observed in the Far Detector, compared to an expected number of $754\pm28\rm{(stat)}\pm{37}\rm{(syst)}$ for oscillations among three active flavors. The fraction $f_s$ of disappearing \numu~that may transition to $\nu_s$ is found to be less than 22\% at the 90\%~C.L.

\end{abstract}
\maketitle

The disappearance of muon neutrinos as they propagate from their production source is well established by experimental evidence accumulated over the past several decades~\cite{ref:previous,ref:superk,Ahn:2006zza,ref:CCPUBS, ref:newCCPUBS}. Although the deficit of \numu~charged-current (CC) interactions is generally interpreted as due to oscillations between the weak flavor states of active neutrinos, with $\num\rightarrow\nut$ transitions representing the dominant channel~\cite{superktau,ref:CCPUBS,ref:newCCPUBS}, more exotic scenarios where active neutrinos oscillate into an unseen sterile neutrino flavor, $\nus$, are not excluded. The possible existence of one or more light sterile neutrinos, in addition to the three active flavors, has been widely discussed~\cite{steriletheory} and could contribute to the understanding of the neutrino mass spectrum~\cite{deGouvea:2006gz} or explain apparent differences in behavior between neutrino and antineutrino oscillations~\cite{ref:Nelson}. Interest in sterile neutrinos has been renewed with the latest observations from antineutrino running in the MiniBooNE experiment, which may be explained by mixing models incorporating one or more sterile neutrinos~\cite{ref:MiniBoone}. Furthermore, recent results from the WMAP experiment might suggest the existence of a fourth neutrino generation with mass less than \unit[0.58]{eV}~\cite{ref:wmap}.

MINOS can probe active to sterile neutrino mixing driven by the atmospheric mass-squared splitting by measuring the rate of  neutral-current (NC) events at two locations, over a baseline of \unit[735]{km}. Because NC cross-sections are identical among the three active flavors, NC event rates are unaffected by standard neutrino mixing. However, oscillations into a sterile noninteracting neutrino flavor would result in an energy-dependent depletion of NC events at the far site. This letter reports results from a search for sterile neutrino mixing, using a data sample twice as large as that used in previous publications~\cite{ref:NCPUBS}. The data are compared to models where these neutrino oscillations are driven either by the atmospheric mass scale $\Delta m_{32}^2$ alone or along with a mass-squared splitting $\Delta m_{43}^2$ having magnitude $\mathcal{O}$(\unit[1]{eV$^{2}$}).

MINOS measures neutrinos from the NuMI beam~\cite{ref:numi} using two detectors: the \unit[980]{ton} (\unit[27]{ton} fiducial) Near Detector (ND), located \unit[1.04]{km} downstream of the beam target at Fermilab; and the \unit[5.4]{kton} (\unit[4.0]{kton} fiducial) Far Detector (FD), placed \unit[735]{km} downstream of the target in the Soudan Underground Laboratory, in Minnesota~\cite{ref:nim}. The detectors are planar steel and scintillator tracking calorimeters. Each plane is composed of \unit[2.54]{cm} thick steel and \unit[1]{cm} thick plastic scintillator arranged in \unit[4.1]{cm} wide strips. The energy resolution function for neutrino-induced hadronic showers is approximately $56\%/\sqrt{E}$~\cite{ref:mike}.  The data were collected in an exposure of $7.07\times10^{20}$~protons on target taken exclusively with a beam configuration for which the peak neutrino event energy is \unit[3.3]{GeV}. The NuMI beam includes a 1.3\% ($\nu_e+\bar{\nu}_e$) contamination primarily from the decay of muons originating in kaon and pion decays.  

In addition to the increased statistics, the new analysis includes changes to the shower reconstruction, whereby active strips with fewer than two photoelectrons of pulse height are excluded from the clustering algorithms, reducing effects from crosstalk. The analysis also benefits from a complete reevaluation of both the event selection criteria and the effects of systematic uncertainties.

The ND registers a high event rate during operation, with multiple neutrino interactions occurring throughout the detector for each beam spill. The total activity recorded during a spill is separated into activity slices using timing and spatial criteria~\cite{ref:CCPUBS}. Ideally, each activity slice would correspond to one neutrino interaction, but some failure modes result in one activity slice containing information from different interactions, for which separate events may then be reconstructed. Simulations show that these failure modes increase the number of interactions selected as NC with reconstructed energy ($E_{\rm{reco}}$) lower than~\unit[1]{GeV} by 37\% in the ND.  This background is reduced to 11\% by removing a reconstructed event if it contains less than half of the total energy deposited in the activity slice, or if the event has fewer than three contiguous planes with at least two photoelectrons read out in each plane~\cite{ref:ThesisGemma}.

Only a few beam-related events are recorded each day in the FD fiducial volume. Interactions are selected for the analysis if they occur between \unit[2]{$\mu\rm{s}$} before and \unit[12]{$\mu\rm{s}$} after the expected start time of the \unit[10]{$\mu\rm{s}$} spill at the FD. Possible backgrounds due to detector noise or cosmic-ray muons in coincidence with the spill window are removed by various selections~\cite{ref:NCPUBS}. The remaining nonbeam backgrounds after application of these criteria represent only 0.5\% of the expected NC interaction rate in the FD.

In the MINOS detectors, NC interactions give rise to events with a short diffuse hadronic shower and either small or no tracks, whereas CC events typically display a long muon track accompanied by hadronic activity at the event vertex. Events crossing fewer than 47 planes for which no track is reconstructed are selected as NC; events crossing fewer than 47 planes that contain a track are classified as NC only if the track extends less than 6 planes beyond the shower~\cite{ref:ThesisGemma}. These selections result in an NC-selected sample with 89\% efficiency and 61\% purity.
Events that fail both NC criteria are selected as CC if they pass the classification procedures used by the MINOS muon neutrino CC disappearance analysis~\cite{ref:newCCPUBS}; otherwise they are removed from the analysis. Highly inelastic \numu~and $\bar{\nu}_\mu$~CC events, where the muon track is not distinguishable from the hadronic shower, are the main source of background for the NC-selected spectrum.

\begin{figure}
  \includegraphics[width=0.95\linewidth]{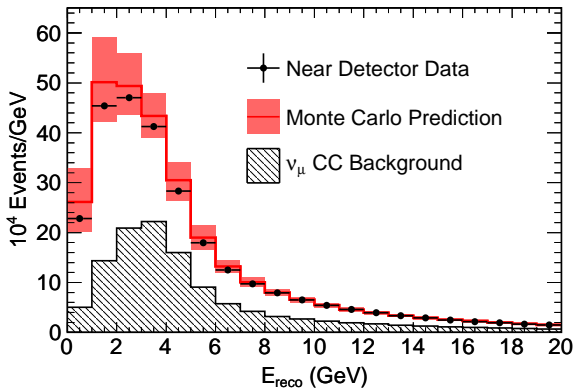}
  \caption{Reconstructed energy spectrum of NC-selected events in the ND (solid points) compared to the Monte Carlo prediction (open histogram) shown with \unit[1]{$\sigma$} systematic errors (shaded band). Also displayed is the simulation of the background from misidentified CC events (hatched histogram).}
  \label{fig:NDSpectrum}
\end{figure}

\begin{figure}
  \includegraphics[width=0.95\linewidth]{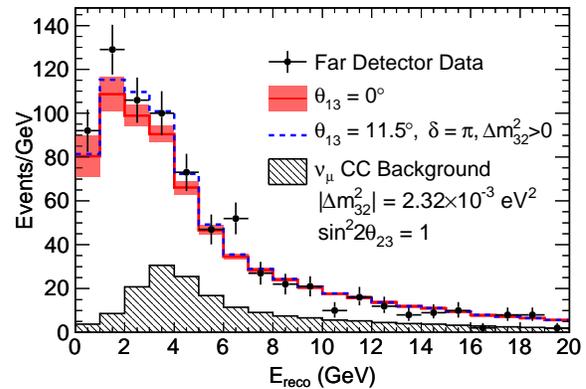}
  \caption{Reconstructed energy spectrum of NC-selected events in the FD (points with statistical errors) compared with predictions for standard three-flavor mixing with and without $\nu_e$~appearance at the MINOS 90\% C.L. limit~\cite{ref:minosnue} (dashed and solid lines respectively).}
  \label{fig:FDSpectrum}
\end{figure}

The predicted NC energy spectrum in the FD is obtained using the ND data. An estimate of the ratio of events in the FD and ND as a function of reconstructed energy, $E_{\rm{reco}}$, is calculated from Monte Carlo simulations. The ratio is multiplied by the observed ND energy spectrum to produce the predicted FD spectrum~\cite{ref:ThesisJason}. Alternative methods of predicting the FD energy spectrum~\cite{ref:ThesisPhil, ref:ThesisRobert} yielded very similar results.  To avoid biases, the analysis selections and procedures were determined prior to examining the FD data. Figures~\ref{fig:NDSpectrum} and~\ref{fig:FDSpectrum} show the reconstructed energy spectra in each detector. The relatively low number of events observed in the  ND energy spectrum for $E_{\rm{reco}}<\unit[1]{GeV}$ is a consequence of the application of the ND-specific selections described above. 

 The analysis classifies 97\% of \nue-induced CC events as NC, requiring the possibility of \nue~appearance to be considered when extracting results.   The  normal neutrino mass hierarchy is assumed with $\theta_{13}=11.5^\circ$ and $\delta_{CP}=\pi$ at the MINOS 90\% C.L. limit~\cite{ref:minosnue}.  Mikheyev-Smirnov-Wolfenstein-like matter effects~\cite{ref:MSW}, due to differences between the matter potentials for active and sterile neutrinos, are at the subpercent level for the MINOS baseline and are neglected in this analysis.

The selection procedures identify 802 NC interaction candidates in the FD, with $754\pm28\rm{(stat)}\pm{37}\rm{(syst)}$ events expected from standard three-flavor mixing (assuming $\theta_{13}=0^\circ$). An excess relative to the $\theta_{13}=0^\circ$ prediction is observed in Fig.~\ref{fig:FDSpectrum} for $1<E_{\rm{reco}}<\unit[5]{\rm{GeV}}$. This excess does not present significant evidence for new neutrino phenomena and is treated as a statistical fluctuation.  The agreement between the observed and predicted NC spectra is quantified using the statistic $R$:
\begin{equation} 
R\equiv\frac{N_{\rm{data}}-B_{\rm{CC}}}{S_{\rm{NC}}},
\end{equation} 
where $N_{\text{data}}$ is the observed number of events, $B_{\text{CC}}$ is the predicted CC background from all flavors, and  $S_{\text{NC}}$ is the expected number of NC interactions. The values of $N_{\text{data}}$, $S_{\text{NC}}$ and contributions to $B_{\text{CC}}$ for various reconstructed energy ranges are shown in Table~\ref{table:nums}.  
The values of $R$ obtained for each energy range show no evidence of a depletion in the NC flux at the FD, supporting the hypothesis that standard three-flavor oscillations explain the data. A value of $R=1.09\pm0.06\text{(stat)}\pm0.05\text{(syst)}-0.08(\nue)$ is measured over the full energy range $0-\unit[120]{GeV}$, where the last term is the change resulting from inclusion of \nue~appearance at its maximally-allowed value from the MINOS 90\% C.L. limit. Therefore, the depletion of the total NC event rate is less than $3.2\%~($11.2\%) at 90\% C.L., where the value in parentheses is obtained assuming \nue~appearance. 
\begin{table} [!htbp]
\begin{tabular}{llcccc}
\hline
$E_{\rm{reco}}$ (GeV)      & $N_{\text{data}}$ & ~~$S_{\text{NC}}$~~ & ~~$B^{\nu_\mu}_{\text{CC}}$~~ & ~~$B^{\nu_\tau}_{\text{CC}}$~~ & ~~$B^{\nu_e}_{\text{CC}}$~~ \bigstrut \\ \hline
$0-1$    & 92 & 76.0 & 3.8 & 0.3 & 0.3~(1.4) \bigstrut  \\
$1-2$    & 129 & 98.0 & 8.6 & 1.1 & 1.0~(7.6) \bigstrut  \\
$2-3$    & 106 & 74.4 & 20.8 & 1.8 & 1.8~(12.6) \bigstrut  \\
$3-4$    & 100 & 55.4 & 30.6 & 2.1 & 2.4~(12.7) \bigstrut  \\
$4-6$    & 120 & 63.1 & 42.3 & 2.9 & 4.4~(13.4) \bigstrut  \\
$6-120$  & 255 & 151.0 & 87.1 & 4.3 & 24.5~(27.8) \bigstrut  \\

\hline
$0-1$ &\multicolumn{5}{l}{$R=1.15\pm0.13\pm0.12-0.01(\nu_e)$} \bigstrut\\
$1-2$ &\multicolumn{5}{l}{$R=1.21\pm0.12\pm0.08-0.07(\nu_e)$} \bigstrut\\
$2-3$ &\multicolumn{5}{l}{$R=1.10\pm0.14\pm0.06-0.15(\nu_e)$} \bigstrut\\
$3-4$ &\multicolumn{5}{l}{$R=1.17\pm0.18\pm0.07-0.19(\nu_e)$} \bigstrut\\
$4-6$ &\multicolumn{5}{l}{$R=1.12\pm0.17\pm0.08-0.15(\nu_e)$} \bigstrut\\
$6-120$ &\multicolumn{5}{l}{$R=0.92\pm0.11\pm0.06-0.02(\nu_e)$} \bigstrut\\
$0-120$ &\multicolumn{5}{l}{$R=1.09\pm0.06\pm0.05-0.08(\nu_e)$} \bigstrut \\
\hline
\end{tabular}\\
\caption{The $R$ statistic and its constituent components for several reconstructed energy ranges.  The numbers shown in parentheses include $\nu_e$~appearance with $\theta_{13}=11.5^\circ$ and $\delta_{CP}=\pi$.  The displayed uncertainties are statistical, systematic, and the uncertainty associated with $\nu_e$~appearance.}  
\label{table:nums}
\end{table}

The data are compared with two models of neutrino oscillations that allow admixture with one sterile neutrino. In the first model, identified as $m_4=m_1$, the first and fourth mass eigenstates are treated as degenerate and the oscillatory behavior is assumed to be driven only by the atmospheric mass scale. The second model, referred to as $m_4\gg m_3$, assumes a large difference between the fourth and third mass eigenstates, introducing an additional mass scale $\Delta m_{43}^2$ with magnitude $\mathcal{O}$(\unit[1]{eV$^{2}$}), so that no oscillation-induced change of the neutrino event rate is measurable at the ND site, but rapid oscillations are predicted at the FD location.  The latter model is also sensitive to portions of the region of interest studied in the MiniBooNE antineutrino data~\cite{ref:MiniBoone2}. Detailed descriptions of these models are provided in Refs.~\cite{ref:NCPUBS} and~\cite{ref:Donini}.


Both the NC-selected energy spectrum shown in Fig.~\ref{fig:FDSpectrum} and the CC-selected spectrum in the FD data are used in the fits to the oscillation models. The dominant systematic uncertainties, discussed below, are added as nuisance parameters to the $\chi^2$ function used in the fits. The best-fit values are summarized in Table~\ref{tab:fit_vals}, which displays results with null and maximal $\nu_e$ appearance. The best=fit value obtained for $|\Delta m^2_{32}|$ is consistent with the results from the muon neutrino CC disappearance analysis~\cite{ref:CCPUBS}. A 90\% C.L. limit on the sterile mixing angle of $\theta_{34} < 26^\circ\,(37^\circ)$ is found for the $m_{4} = m_{1}$ model. For the $m_{4}\gg m_{3}$ model, limits of $\theta_{24} < 7^\circ\,(8^\circ)$ and $\theta_{34} < 26^\circ\,(37^\circ)$ are obtained at the 90\% C.L.~\cite{ref:CDHS}. Pure $\nu_\mu \rightarrow\nu_s$ oscillations are excluded at 99.8\%~(96.2\%)~C.L. The numbers in parentheses represent the limits extracted for maximal $\nue$~appearance.

The coupling between active and sterile neutrinos may also be quantified in terms of the fraction of disappearing \numu~that oscillate into $\nus$, $f_{s}~\equiv~P_{\nu_\mu\rightarrow\nu_s}/(1-P_{\nu_\mu\rightarrow\nu_\mu})$, where the $P_{\nu_\mu\rightarrow\nu_x}$ refer to neutrino oscillation probabilities.
For both sterile oscillation models, the limit set on the disappearance fraction is  $f_{s} < 0.22\,(0.40)$ at the 90\% C.L. This limit represents a 57\% improvement on the previous limit set by MINOS if \nue~appearance is neglected~\cite{ref:NCPUBS}, achieved from increased statistics and the reduction of systematic uncertainties in this analysis. 
\begin{table}[!h]
  \begin{tabular}{|c*{6}{|c}|}
\hline
    Model & $\theta_{13}$   \bigstrut & $\chi^2$/D.O.F. & $\theta_{23}$  & $\theta_{24}$ & $\theta_{34}$ & $f_{s}$\\
    \hline
    \multirow{2}{*}{$m_4 = m_1$} & $0$   \bigstrut & 130.4/123 & $45.0^{+7}_{-7}$ & - & $0.0^{+17}_{-0.0}$  & 0.22\\
                                     & $11.5$ \bigstrut & 128.5/123 & $45.6^{+7}_{-7}$ & - & $0.0^{+25}_{-0.0}$ & 0.40\\
    \hline
    \multirow{2}{*}{$m_4\gg m_3$} & $0$   \bigstrut & 130.4/122 & ${45.0}^{+7}_{-7}$ & ${0.0}^{+5}_{-0.0}$ & ${0.0}^{+17}_{-0.0}$ & 0.22\\
                                 & $11.5$ \bigstrut & 128.5/122 & ${45.6}^{+7}_{-7}$ & ${0.0}^{+5}_{-0.0}$ & ${0.0}^{+25}_{-0.0}$ & 0.40\\
    \hline
  \end{tabular}
  \caption{Best-fit values and uncertainty ranges in degrees for the angles of the two neutrino oscillation scenarios including a sterile neutrino. The results shown assume either no \nue~appearance or \nue~appearance at the MINOS 90\% C.L. limit. The quantity $f_{s}$ is the maximum allowed fraction of disappearing $\nu_{\mu}$ that may transition to $\nu_{s}$.}
  \label{tab:fit_vals}
\end{table}

The dominant systematic uncertainties were reevaluated, in some cases with more sensitive methodologies, leading to reductions from the previous analysis. 
Visual scanning techniques were used to asses the reconstruction algorithms allowing a reduction of the uncertainty due to the relative normalization between the two detectors from 4\% to 2.2\%~\cite{ref:ThesisPhil}.  The absolute scale of the hadronic energy contributes a maximum of 10\% to the uncertainty for $E_{\rm{reco}}\le\unit[0.5]{GeV}$ to a minimum of 6.5\% when $E_{\rm{reco}}>\unit[10]{GeV}$. This new treatment combines a constant 5.6\% uncertainty in the detector response to single hadrons with an energy-dependent uncertainty due to hadronization model and intranuclear effects~\cite{ref:abshadsyst}. The uncertainty on the relative energy scale between detectors, computed from interdetector calibration studies, is 2.1\%.  The uncertainty due to ND-specific selections, determined by varying each selection criterion to assess its effects in the ND energy spectrum, is 10\% for $E_{\rm{reco}}\le\unit[1]{GeV}$, between 4.8\% and 2.1\% for $0.5<E_{\rm{reco}}\le\unit[2.5]{GeV}$, and negligible for higher energies~\cite{ref:ThesisGemma}.  The uncertainties stemming from FD-specific selections, evaluated in the same manner as the ND-specific ones, are calculated to be 5\% for $E_{\rm{reco}}\le\unit[0.5]{GeV}$, and between 2.5\% and 1\% for $0.5<E_{\rm{reco}}\le\unit[120]{GeV}$. The uncertainty in the size of the \numu-CC background, computed using ND data acquired in different beam configurations~\cite{ref:NCPUBS, ref:ThesisPhil}, remains 15\%.

The effects of other systematic uncertainties, such as those due to the physics of neutrino interactions, largely cancel in the FD prediction due to the similarity in materials and planar separation of the ND and FD. These additional uncertainties are the principal contributors to the error band shown in the ND energy spectrum in Fig.~\ref{fig:NDSpectrum}, but their effects on the FD error band shown in Fig.~\ref{fig:FDSpectrum} are negligible. Table~\ref{tab:systs} summarizes the variation in best-fit values obtained for the oscillation parameters when the effects of the dominant systematic uncertainties are applied to the ND and FD energy spectra but not included in the fit. The largest systematic-induced shift in the value of {\it R} over the full energy range is 2.9\% and is due to the relative normalization uncertainty. With the exception of the energy scale uncertainties, which do not modify the total number of events and thus have no effect on {\it R}, the other dominant systematic uncertainties each induce shifts of 2.3\%.  
\begin{table}
\begin{tabular}{|c||c|c||c|c|c|}
\hline
\multirow{2}{*}{Uncertainty} \bigstrut 
&\multicolumn{2}{|c||}{$m_4 = m_1$} \bigstrut 
& \multicolumn{3}{c|}{$m_4\gg m_3$} \bigstrut \\
\cline{2-6}
& $\Delta(\theta_{23})$ & $\Delta(\theta_{34})$ \bigstrut
& $\Delta(\theta_{23})$ & $\Delta(\theta_{24})$ \bigstrut
& $\Delta(\theta_{34})$ \\
\hline
Absolute $E_{\rm{Hadronic}}$\bigstrut & 0.2   & 6.1    & 0.2  & 0.8   & 9.5  \\
Relative $E_{\rm{Hadronic}}$\bigstrut & 0.3   & 5.9    & 0.4   &  1.2  &  9.4  \\
Normalization                  \bigstrut & 0.2   & 11.2   & 0.0   &  4.2  &  6.7  \\
CC Background                  \bigstrut & 0.1   & 12.1   & 0.2   &  0.3  &  10.0  \\
ND Selection                   \bigstrut & 0.2   & 15.1   & 0.4   &  0.7  &  13.8 \\
FD Selection                   \bigstrut & 0.1   & 12.5   & 0.1   &  0.7  &  7.4 \\
\hline
Total                          \bigstrut & 0.5   & 27.0   & 0.6   &  4.6  &  23.8 \\
\hline
\end{tabular}
\caption{Magnitude in degrees of the mixing-angle deviations introduced by the major systematic uncertainties from best-fit results in which systematic shifts have been neglected.}
\label{tab:systs}
\end{table}



In summary, new results are presented from a search for active to sterile neutrino mixing, based upon a data sample with double the event statistics of previous MINOS analyses. A total of 802 NC event candidates is observed in the FD data, compared to $754\pm28\rm{(stat)}\pm{37}\rm{(syst)}$ events expected from standard oscillations. Therefore, no evidence for depletion of NC events is observed in the FD at a distance of \unit[735]{km} from the production target. The most stringent constraint to date is placed on the fraction of active neutrinos that transition to sterile neutrinos, $f_{s} < 0.22\,(0.40)$ at the 90\%~C.L., where the number in parentheses denotes the limit assuming \nue~appearance. The results support the hypothesis that \numu~disappearance observed in MINOS is dominated by oscillations among active neutrino species.

This work was supported by the US DOE; the UK STFC; the US NSF; the State and University of Minnesota; the University of Athens, Greece; and Brazil's FAPESP, CNPq, and CAPES. We gratefully acknowledge the Minnesota DNR, the crew of the Soudan Underground Laboratory, and the personnel of Fermilab for their contribution to this effort.


\begin{thebibliography}{24}
\expandafter\ifx\csname natexlab\endcsname\relax\def\natexlab#1{#1}\fi
\expandafter\ifx\csname bibnamefont\endcsname\relax
  \def\bibnamefont#1{#1}\fi
\expandafter\ifx\csname bibfnamefont\endcsname\relax
  \def\bibfnamefont#1{#1}\fi
\expandafter\ifx\csname citenamefont\endcsname\relax
  \def\citenamefont#1{#1}\fi
\expandafter\ifx\csname url\endcsname\relax
  \def\url#1{\texttt{#1}}\fi
\expandafter\ifx\csname urlprefix\endcsname\relax\def\urlprefix{URL }\fi
\providecommand{\bibinfo}[2]{#2}
\providecommand{\eprint}[2][]{\url{#2}}

\bibitem[{ref({\natexlab{a}})}]{ref:previous}
\bibinfo{note}{R.~Becker-Szendy et al. (IMB-3), Phys. Rev. D {\bf 46}, 3720
  (1992); K.~S.~Hirata et al. (Kamiokande), Phys. Lett. B {\bf 280}, 146
  (1992); W.W.M.~Allison et al. (Soudan-2), Phys. Rev. D {\bf 72}, 052005
  (2005); M.~Ambrosio et al. (MACRO), Eur. Phys. J. C. {\bf 36}, 323 (2004)}.

\bibitem[{ref({\natexlab{b}})}]{ref:superk}
\bibinfo{note}{Y.~Fukuda et al. (Super-Kamiokande), Phys. Rev. Lett. {\bf 81},
  1562 (1998); Y.~Ashie et al. (Super-Kamiokande), Phys. Rev. Lett. {\bf 93},
  101801 (2004); Y.~Ashie et al. (Super-Kamiokande), Phys. Rev. D {\bf 71},
  112005 (2005)}.

\bibitem[{\citenamefont{Ahn et~al.}(2006)}]{Ahn:2006zza}
\bibinfo{author}{\bibfnamefont{M.~H.} \bibnamefont{Ahn}} \bibnamefont{et~al.}
  (\bibinfo{collaboration}{K2K}), \bibinfo{journal}{Phys. Rev. D}
  \textbf{\bibinfo{volume}{74}}, \bibinfo{pages}{072003}
  (\bibinfo{year}{2006}).

\bibitem[{ref({\natexlab{c}})}]{ref:CCPUBS}
\bibinfo{note}{P.~Adamson et al. (MINOS), Phys. Rev. Lett. {\bf 101} 131802
  (2008); P.~Adamson et al. (MINOS), Phys. Rev. D {\bf 77} 072002 (2008);
  D.~G.~Michael et al. (MINOS), Phys. Rev. Lett. {\bf 97}, 191801 (2006)}.

\bibitem[{ref({\natexlab{d}})}]{ref:newCCPUBS}
\bibinfo{note}{P.~Adamson et al. (MINOS), Phys. Rev. Lett. {\bf 106} 181801
  (2011)}.

\bibitem[{sup()}]{superktau}
\bibinfo{note}{Y.~Fukuda et al. (Super-Kamiokande), Phys. Rev. Lett. {\bf 85},
  3999 (2000); K.~Abe et al. (Super-Kamiokande), Phys. Rev. Lett. {\bf 97},
  171801 (2006); W.~Wang, Ph.D. Thesis, Boston University (2007)}.

\bibitem[{ste()}]{steriletheory}
\bibinfo{note}{G.~L.~Fogli, E.~Lisi and A.~Marrone, Phys. Rev. D {\bf 64}, 093005 (2001);
  A.~Donini et al. JHEP {\bf 12}, 013 (2007); A.~Dighe and S.~Ray, Phys. Rev. D
  {\bf 76}, 113001 (2007)}.

\bibitem[{\citenamefont{de~Gouvea et~al.}(2007)\citenamefont{de~Gouvea,
  Jenkins, and Vasudevan}}]{deGouvea:2006gz}
\bibinfo{author}{\bibfnamefont{A.}~\bibnamefont{de~Gouvea}},
  \bibinfo{author}{\bibfnamefont{J.}~\bibnamefont{Jenkins}}, \bibnamefont{and}
  \bibinfo{author}{\bibfnamefont{N.}~\bibnamefont{Vasudevan}},
  \bibinfo{journal}{Phys. Rev. D} \textbf{\bibinfo{volume}{75}},
  \bibinfo{pages}{013003} (\bibinfo{year}{2007}).

\bibitem[{\citenamefont{Engelhardt et~al.}(2010)\citenamefont{Engelhardt,
  Nelson, and Walsh}}]{ref:Nelson}
\bibinfo{author}{\bibfnamefont{N.}~\bibnamefont{Engelhardt}},
  \bibinfo{author}{\bibfnamefont{A.~E.} \bibnamefont{Nelson}},
  \bibnamefont{and} \bibinfo{author}{\bibfnamefont{J.~R.} \bibnamefont{Walsh}},
  \bibinfo{journal}{Phys. Rev. D} \textbf{\bibinfo{volume}{81}},
  \bibinfo{pages}{113001} (\bibinfo{year}{2010}).

\bibitem[{\citenamefont{Karagiorgi et~al.}(2009)\citenamefont{Karagiorgi,
  Djurcic, Conrad, Shaevitz, and Sorel}}]{ref:MiniBoone}
\bibinfo{author}{\bibfnamefont{G.}~\bibnamefont{Karagiorgi}},
  \bibinfo{author}{\bibfnamefont{Z.}~\bibnamefont{Djurcic}},
  \bibinfo{author}{\bibfnamefont{J.~M.} \bibnamefont{Conrad}},
  \bibinfo{author}{\bibfnamefont{M.~H.} \bibnamefont{Shaevitz}},
  \bibnamefont{and} \bibinfo{author}{\bibfnamefont{M.}~\bibnamefont{Sorel}},
  \bibinfo{journal}{Phys. Rev. D} \textbf{\bibinfo{volume}{80}},
  \bibinfo{pages}{073001} (\bibinfo{year}{2009}).

\bibitem[{ref({\natexlab{e}})}]{ref:wmap}
\bibinfo{note}{E.~Komatsu et al. (WMAP), Astrophys. J. Suppl. Ser. {\bf 192}, 18 (2011);
  J.~Hamann, S.~Hannestad, G.~G.~Raffelt, I.~Tamborra and Y.~Y.~Y.~Wong, Phys
  Rev. Lett. {\bf 105}, 181301 (2010).}

\bibitem[{ref({\natexlab{f}})}]{ref:NCPUBS}
\bibinfo{note}{P.~Adamson et al. (MINOS), Phys. Rev. D {\bf 81} 052004 (2010);
  P.~Adamson et al. (MINOS), Phys. Rev. Lett. {\bf 101}, 221804 (2008)}.

\bibitem[{ref({\natexlab{g}})}]{ref:numi}
\bibinfo{note}{S. Kopp, in Proc. 2005 IEEE Part. Accel. Conf., 2005, arXiv:physics/0508001.}

\bibitem[{\citenamefont{Michael et~al.}(2008)}]{ref:nim}
\bibinfo{author}{\bibfnamefont{D.~G.} \bibnamefont{Michael}}
  \bibnamefont{et~al.} (\bibinfo{collaboration}{MINOS}),
  \bibinfo{journal}{Nucl. Instrum. Meth. A} \textbf{\bibinfo{volume}{596}},
  \bibinfo{pages}{190} (\bibinfo{year}{2008}).

\bibitem[{\citenamefont{Kordosky}(2004)}]{ref:mike}
\bibinfo{author}{\bibfnamefont{M.}~\bibnamefont{Kordosky}},
  \bibinfo{journal}{Ph.D. Thesis, University of Texas at Austin}
  (\bibinfo{year}{2004}).

\bibitem[{\citenamefont{Tinti}(2010)}]{ref:ThesisGemma}
\bibinfo{author}{\bibfnamefont{G.}~\bibnamefont{Tinti}},
  \bibinfo{journal}{D.Phil. Thesis, Oxford University}  (\bibinfo{year}{2010}).

\bibitem[{\citenamefont{Koskinen}(2009)}]{ref:ThesisJason}
\bibinfo{author}{\bibfnamefont{D.~J.} \bibnamefont{Koskinen}},
  \bibinfo{journal}{Ph.D. Thesis, University College London}
  (\bibinfo{year}{2009}).

\bibitem[{\citenamefont{Rodrigues}(2010)}]{ref:ThesisPhil}
\bibinfo{author}{\bibfnamefont{P.~A.} \bibnamefont{Rodrigues}},
  \bibinfo{journal}{D.Phil. Thesis, Oxford University}  (\bibinfo{year}{2010}).

\bibitem[{\citenamefont{Pittam}(2010)}]{ref:ThesisRobert}
\bibinfo{author}{\bibfnamefont{R.}~\bibnamefont{Pittam}},
  \bibinfo{journal}{D.Phil. Thesis, Oxford University}  (\bibinfo{year}{2010}).

\bibitem[{\citenamefont{Adamson et~al.}(2010)}]{ref:minosnue}
\bibinfo{author}{\bibfnamefont{P.}~\bibnamefont{Adamson}} \bibnamefont{et~al.}
  (\bibinfo{collaboration}{MINOS}), \bibinfo{journal}{Phys. Rev. D}
  \textbf{\bibinfo{volume}{82}}, \bibinfo{pages}{051102(R)}
  (\bibinfo{year}{2010}).

\bibitem[{ref({\natexlab{h}})}]{ref:MSW}
\bibinfo{note}{L.~Wolfenstein, Phys. Rev. D {\bf 17}, 2369 (1978);
  S.~P.~Mikheyev and A.~Yu.~Smirnov, Sov. J. Nucl. Phys. {\bf 42}, 913 (1985);
  see, for example, P.~Lipari and M.~Lusignoli, Phys. Rev. D {\bf 58}, 073005
  (1998)}.

\bibitem[{\citenamefont{Aguilar-Arevalo et~al.}(2010)}]{ref:MiniBoone2}
\bibinfo{author}{\bibfnamefont{A.~A.} \bibnamefont{Aguilar-Arevalo}}
  \bibnamefont{et~al.} (\bibinfo{collaboration}{MiniBooNE}),
  \bibinfo{journal}{Phys. Rev. Lett.} \textbf{\bibinfo{volume}{105}},
  \bibinfo{pages}{181801} (\bibinfo{year}{2010}).

\bibitem[{ref({\natexlab{i}})}]{ref:Donini}
\bibinfo{note}{A. Donini, M. B. Gavela, P. Hernandez and S. Rigolin, Nucl.
  Phys. B {\bf 574} 23-42 (2000); A. Donini, M. Maltoni, D. Meloni, P.
  Migliozzi and F. Terranova, J. High En. Phys. {\bf 12} 013 (2007)}.

\bibitem[{ref({\natexlab{j}})}]{ref:CDHS}
\bibinfo{note}{F. Dydak et al. (CDHS), Phys. Lett. B {\bf 134}, 281 (1984); I.
  Stockdale et al. (CCFR), Z. Phys. C {\bf 27}, 53 (1985); Previous experiments
  looking for \numu~disappearance driven by $\Delta m^2\sim\unit[1]{eV^2}$ did
  so assuming that it depends only on $\theta_{24}$. The 90\% C.L. excluded
  range in $\sin^2(2\theta_{24})$ is expanded in this analysis to 0.06-1.0 over
  the mass-squared range $\unit[0.3-2.5]{eV^2}$. MINOS is the only experiment
  to limit the size of $\theta_{34}$}.

\bibitem[{\citenamefont{Dytman et~al.}(2008)\citenamefont{Dytman, Gallagher,
  and Kordosky}}]{ref:abshadsyst}
\bibinfo{author}{\bibfnamefont{S.}~\bibnamefont{Dytman}},
  \bibinfo{author}{\bibfnamefont{H.}~\bibnamefont{Gallagher}},
  \bibnamefont{and} \bibinfo{author}{\bibfnamefont{M.}~\bibnamefont{Kordosky}}
  (\bibinfo{year}{2008}), \eprint{arXiv:hep-ex/0806.2119}.

\end{thebibliography}
\end{document}